\documentclass[12pt]{article}

\usepackage{amsfonts,dsfont}
\usepackage{hyperref}
\hypersetup{
	colorlinks = true,
	citecolor = cyan,
	linkcolor = magenta,
	linktocpage
}

\usepackage[dvipsnames]{xcolor}
\usepackage{amsmath, amssymb}
\usepackage{amsthm}
\usepackage{bm}
\usepackage{color}
\usepackage{comment}
\usepackage{enumerate}
\usepackage{geometry}
\usepackage{mathtools}
\usepackage{tikz}
\usepackage{tikz-cd}
\allowdisplaybreaks

\let\oldhref\href
\renewcommand{\href}[2]{\oldhref{#1}{\texttt{#2}}}

\usepackage{fancyhdr}
\fancypagestyle{firstpage}{
    \fancyhf{} 
    \fancyfoot[C]{\thepage}
    \fancyfoot[R]{\today} 
}

\theoremstyle{definition}\newtheorem{definition}{Definition}
\theoremstyle{definition}
\theoremstyle{definition}\newtheorem{remark}[definition]{Remark}
\theoremstyle{definition}
\theoremstyle{definition}
\theoremstyle{definition}
\theoremstyle{definition}\newtheorem{example}[definition]{Example}
\theoremstyle{definition}
\theoremstyle{definition}
\theoremstyle{definition}
\theoremstyle{definition}

\counterwithin{equation}{section}

\newcommand{\D}{\mathcal{D}}
\newcommand{\G}{\mathcal{G}}

\newcommand{\cI}{\mathcal{I}}

\newcommand{\R}{\mathbb{R}}

\newcommand{\Dt}{\mathrm{D}}

\newcommand{\ii}{^{(\infty)}}

\newcommand{\n}{^{(n)}}

\newcommand{\indK}{\cI_{\K}}
\newcommand{\indKn}{\indK^{\leq n}}
\newcommand{\indKa}{\indK^\alpha }

\newcommand{\A}{\mathcal{A}}
\newcommand{\mB}{\mathcal{B}}

\newcommand{\mS}{\mathcal{S}}

\newcommand{\mI}{\mathcal{I}}
\newcommand{\K}{\mathcal{K}}

\newcommand{\mV}{\mathcal{V}}
\newcommand{\mX}{\mathcal{X}}

\def\r#1{\overline{#1}{}}

\newcommand{\rG}{\r{\G}}
\newcommand{\rGn}{\rG\n}

\newcommand{\rU}{\r{U}}

\newcommand{\rX}{\r{X}}

\newcommand{\rY}{\r{Y}}
\newcommand{\rZ}{\r{Z}}

\newcommand{\rrho}{\r{\rho}}

\newcommand{\hU}{\widehat{U}}

\newcommand{\tU}{\widetilde{U}}

\newcommand{\dt}{\mathrm{d}}

\def\ostrut#1#2{\hbox{\vrule height #1pt depth #2pt width 0pt}}

\def\ie{i.e., }

\def\at#1{|_{#1}}
\def\nth{$n$-th }

\def\dom{\text{dom }}
\def\comp{\raise 1pt \hbox{$\,\scriptstyle\circ\,$}}

\def\ps#1{^{(#1)}}
\def\ro#1{{\rm #1}}
\def\J#1{\ro J{}^{#1}} \def\Jn{\J n}   
\def\j#1{{\rm j}_{#1}} \def\jn{\j n}

\def\Dn{\D^{(n)}}

\def\Gn{\G^{(n)}}

\newcommand{\cls}{\text{cls }}
\newcommand{\prin}{\text{prin}}
\newcommand{\para}{\text{par}}

\def\cone{\mathcal{C}}

\newcommand{\tY}{\widetilde{Y}}

\def\ostrut#1#2{\hbox{\vrule height #1pt depth #2pt width 0pt}}

\begin{document}
\begin{center}
{\Large {\bf Moving Frames for Lie Pseudo-groups: \ostrut0{10}\\A Recursive Implementation}}
\end{center}
\vspace{0.5cm}
\hfill
\begin{minipage}{0.4\linewidth}
Peter J.\ Olver \\
School of Mathematics\\
University of Minnesota \\
Minneapolis, MN\quad 55455\\
\href{mailto:olver@umn.edu}{{\tt olver@umn.edu}}
\end{minipage}
\hfill
\begin{minipage}{0.4\linewidth}
Francis Valiquette\\
Department of Mathematics\\
Monmouth University\\
West Long Branch, NJ\quad 07764\\
\href{mailto:fvalique@monmouth.edu}{{\tt fvalique@monmouth.edu}}
\end{minipage}
\hfill\\[1cm]
\begin{abstract}
A new recursive implementation of the equivariant moving frame method for infinite-dimensional Lie pseudo-group actions is presented.  It allows pseudo-group normalizations to be performed before the prolonged action is computed, thereby avoiding the introduction of unnecessarily complicated intermediate expressions.  Several examples are provided to illustrate the algorithm.
\end{abstract}



\section{Introduction}

\thispagestyle{firstpage} 

The use of moving frames can be traced back to the German mathematician Martin Bartels, who attached a trihedron to each point of a space curve and 
obtained formulas for its infinitesimal displacement which are equivalent to the Frenet--Serret formulas, \cite{AR-1993,L-1997}.  Building of the works of Cotton, \cite{C-1905}, Darboux, \cite{D-1914}, Frenet, \cite{F-1852} Ribaucour, \cite{R-1884} and Serret, \cite{S-1851}, \'Elie Cartan developed in \cite{C-1935,C-1937} the method of moving frames into a powerful procedure for analyzing geometric properties of submanifolds and their invariants under the action of transformation groups.  

In the 1970's and 1980's, several researchers worked on framing Cartan's  constructions within a firm theoretical foundation, \cite{C-1985,G-1978,G-1974,J-1977}, ultimately leading to the modern expositions \cite{BCGGG-1991,C-2017,O-1995}.  At the turn of the twenty-first century, a new formulation of Cartan's method of moving frame emerged, first for finite-dimensional group actions, \cite{FO-1999}, and then for infinite-dimensional Lie pseudo-groups, \cite{OP-2008}.  In the new framework, moving frames are no longer constrained by frame bundles or connections. Instead, given a local (pseudo-)group of transformations of a manifold $M$, a moving frame is defined as an \emph{equivariant map} from the jet bundle of submanifolds $S\subset M$ into the (pseudo-)group. Such an equivariant moving frame sends submanifolds to their normal form power series, whose non-constant Taylor coefficients provide a complete set of differential invariants.  This new formulation has resulted in a wide range of applications, including classical invariant theory, \cite{O-1999},  differential equations, \cite{M-2010,TV-2015,V-2018}, integrable systems, \cite{BV-2020,MO-2010}, the calculus of variations, \cite{GM-2016,KO-2003,TV-2011}, differential geometry, \cite{AV-2021,HO-2007,OSV-2023}, difference equations, \cite{BV-2017,BV-2018,TV-2018}, differential-difference equations, \cite{WH-2024}, numerical analysis, \cite{O-2001a,RV-2013}, object recognition in 2D and 3D images, \cite{FKK-2010}, automatic assembly of jigsaw puzzles, \cite{GOOSST-2016,HO-2014}, machine learning, \cite{O-2026,SBVA-2023}, and much more.  For an overview of the equivariant approach to moving frames and some of its applications, we refer the reader to \cite{M-2010,O-2012}.

The construction of an equivariant moving frame first requires the action to be prolonged to a sufficiently high order submanifold jet space so that the 
prolonged action becomes free. (In this paper we do not consider other types of prolongation, such as to Cartesian products of jet spaces, \cite{O-2001}, multi-space, \cite{O-2001a}, etc.) Provided the prolonged action is also regular, the choice of a cross-section then yields a (right) moving frame by finding the transformation that sends a submanifold jet onto the cross-section.  

Although the idea of prolonging the action until it becomes free is crucial in the development of the general theory, the calculation of the prolonged action, which is based on implicit differentiation, can produce expressions that rapidly overwhelm symbolic software, thereby limiting the practical scope of the method. For this reason, inductive and recursive implementations of the equivariant moving frame method have been developed in \cite{K-2001,K-2003,O-2011,OV-2018,V-2013}.  The inductive algorithm assumes the construction of a moving frame for a subgroup of transformations and uses these data to construct a moving frame for the larger group.  In the recursive implementation, normalizations are implemented at lower order before expressions for the higher order prolonged action is computed.  The recursive algorithms presented in \cite{O-2011,OV-2018} make use of the recurrence relations, \cite{FO-1999,OP-2008}, and the formulas for the Maurer--Cartan forms, \cite{OP-2005}.  

In the present paper, we propose a new recursive algorithm that does not rely on the recurrence relations and the Maurer--Cartan forms.  We will primarily focus our attention on infinite-dimensional Lie pseudo-groups, as our algorithm is the most effective in this setting.  The algorithm is inspired by recent results obtained in \cite{OSV-2025,OV-2026}, where we proved the convergence of normal form power series constructed using the method of equivariant moving frames under mild assumptions on the pseudo-group action and the cross-section defining the moving frame.  One key idea, which is natural when considering the action of a Lie pseudo-group on submanifolds, is the introduction of the notion of reduced Lie pseudo-group, where all pseudo-group transformations are restricted to a prescribed submanifold $S$.  These transformations were shown in \cite{OSV-2025} to satisfy an involutive system of differential equations called the \emph{reduced determining equations}.  To obtain a specific reduced pseudo-group transformation, the reduced determining equations must then be supplemented with well-posed initial conditions. In the present paper, we use the equation that relates the prescribed submanifold $S$ to a normal form to obtain suitable initial conditions for the reduced determining equations.  These initial conditions prescribe a unique pseudo-group transformation, corresponding to the moving frame that sends $S$ to its normal form.  The most important aspect of our construction is that the order in which these initial conditions are imposed on the pseudo-group transformation and when the prolonged action is computed does not matter.  This allows us to perform pseudo-group normalizations at low order before computing the prolonged action, thereby keeping the expressions manageable.

The paper is structured as follows. In Section \ref{sec: pg} we recall the notion of a Lie pseudo-group and introduce the concept of a reduced Lie pseudo-group.  In Section \ref{sec: mf} we recall the standard implementation of the equivariant moving frame method.  Section \ref{sec: rmfi} contains the main contribution of the paper in which our recursive moving frame implementation is introduced.  The concepts are illustrated throughout the paper with a running example, and in Section \ref{sec: ex} more examples are provided.

\section{Lie pseudo-groups and their action of submanifolds}\label{sec: pg}

Let $M$ be an $m$-dimensional analytic manifold, known as the \emph{total space} and let $\D = \D(M)$ denote the Lie pseudo-group of all local analytic diffeomorphisms $\varphi\colon M\to M$, where the notation allows $\varphi$ to be defined only on an open subset of $M$.  We adopt Cartan's notational convention, where lowercase letters denote source coordinates and the corresponding capital letters the target coordinates.  Therefore, given a local diffeomorphism $\varphi \in \D$, its local coordinate formula is $Z=\varphi(z)$, so that the target coordinates $Z = (Z^1,\ldots, Z^m)$ are functions of the source coordinates $z = (z^1,\ldots,z^m)$. Given $0 \leq n < \infty$, let $\Dn \subset \Jn(M,M)$ be the subbundle consisting of all \nth order jets of local diffeomorphisms of $M$.  Local coordinates on $\Dn$ are given by $(z,Z\n)$, where $Z\n = \jn\varphi\at{z}$ collects all the partial derivatives $\partial^k\varphi^a/\partial z^A$ of order $0\leq |A| = k \leq n$ at the source $z \in M$.  We employ symmetric multi-index notation for partial derivatives throughout, so here $A = (a_1,\ldots,a_k)$ with $1 \leq a_k \leq m$.

Let $\G\subset \D$ be a regular Lie pseudo-group action on $M$.  Regularity means that for all sufficiently large $n \gg 0$, the pseudo-group jets $\G\n \subset \D\n$ form a subbundle, and the induced projection $\pi^{n+1}_n \colon \G\ps{n+1}\to \G\n$ forms a fibration.  Furthermore, for $\G$ to be a Lie pseudo-group, every local diffeomorphism $\varphi \in \D$ that satisfies $\jn \varphi \subset \G\n$ for $n\gg 0$ must belong to the pseudo-group, \ie $\varphi \in \G$.  Locally,  $\Gn$ is described by its \nth order determining equations 
\begin{equation}\label{Gn}
\Gn = \big\{\Delta(z,Z\n) = 0\big\},
\end{equation}
whose solutions are the pseudo-group transformations.  The equations \eqref{Gn} are assumed to be involutive, and we refer the reader to \cite{S-2010} for a complete and modern presentation of the theory of involutivity.

\begin{example}
As our running example, we will consider the Lie pseudo-group
\begin{equation}\label{pg1}
X = f(x),\qquad 
Y = f_x(x)\, y + g(x),\qquad
U = u + \frac{f_{xx}(x)\, y + g_x(x)}{f_x(x)},
\end{equation}
where $f\in \D(\R)$, so $f_x(x) \ne 0$, and $g\in C^\infty(\R)$.  (We will use subscripts to indicate derivatives throughout.) The transformations \eqref{pg1} are solutions to the involutive determining equations
\begin{gather*}
X_y = X_u = 0,\qquad  Y_x = (U-u)X_x,\qquad Y_y = X_x,\qquad Y_u=0,\qquad U_u = 1,\\
X_{xx} = U_y X_x,\quad X_{xy} =X_{xu} = X_{yy} = X_{yu} = X_{uu} = 0,\quad
Y_{xx} = \big[U_x + (U-u)U_y\big]X_x,\\
Y_{xy} = U_y X_x,\quad Y_{xu} = Y_{yy}= Y_{yu} = Y_{uu} = 0,\qquad
U_{xu} = U_{yy} = U_{yu}=U_{uu} = 0.
\end{gather*}
\end{example}

Given a Lie pseudo-group $\G$, we are interested in its induced action on $p$-dimensional submanifolds $S \subset M$, where $1\leq p < m = \dim M$.  Locally, we assume that the submanifolds $S$ are local sections of the fiber bundle $\pi\colon M\to \mX$ with $\dim \mX = p$. We introduce the local coordinates $z = (x,u) = (x^1,\ldots,x^p,u^1,\ldots,u^q)$ on the total space $M$, where $p+q = m = \dim M$, so that submanifolds $S$ are given locally as graphs of functions $u=u(x)$.  The corresponding target coordinates of a local diffeomorphism are given by $Z = (X,U) = (X^1,\ldots,X^p,U^1,\ldots,U^q)$.  We then let $\Jn$ denote the  \nth order submanifold jet space, cf.~\cite{O-1995}, with local coordinates given by $z\n = (x,u\n) = (\; \ldots x^i\; \ldots\; u^\alpha_J\; \ldots\;)$ for $i=1,\ldots,p$, $\alpha = 1,\ldots,q$ and $|J| \leq n$.

For a fixed $p$-dimensional submanifold $S \subset M$ given, in local coordinates, by the section $s=\{(x,u(x))\}$, we now restrict every local diffeomorphism $\varphi \in \G$ with $s\subset \dom \varphi$ to the submanifold $S$, so that $\varphi\at{s}$ has local coordinate expression
\[
\r{\varphi}(x)  = \varphi(x,u(x)).
\]
In general, we will use bars to denote the restriction of functions and maps to a prescribed submanifold. This results in what we call the \emph{reduced Lie pseudo-group} $\rG$, in accord with \cite{OSV-2025}.  As with the original Lie pseudo-group $\G$, elements $\r{\varphi}$ of the reduced Lie pseudo-group are solutions to a system of partial differential equations  
\begin{equation}\label{rGn}
\rGn = \big\{\r{\Delta}(z\n,\rZ\n) = 0\big\},
\end{equation}
called the \nth order \emph{reduced determining equations}.  The explicit construction of the  reduced determining equations is based on implicitization, and is explained in detail in \cite{OSV-2025}.  In applications, the determining equations \eqref{rGn} are solved for as many reduced pseudo-group jets as possible; these are called \emph{principal derivatives} and labeled as $\rZ\n_\prin$, while the remaining reduced pseudo-group parameters $\rZ\n_\para$ are called \emph{parametric derivatives}, and serve to parametrize $\rG\n$.

\begin{definition}
A pseudo-group $\G$ is said to be order \emph{$n$ reducible} on the local section $s\colon \mX \to M$ if, for all $x\in \dom s$ with $z\n = \jn s\at{x}$, there is a one-to-one correspondence between the fibers of $\G\n\at{z}$ and $\rGn\at{z\n}$, and hence the fiber dimensions are equal.  The pseudo-group $\G$ is \emph{reducible} on $s$ if it is reducible for all sufficiently large $n\gg 0$.
\end{definition}

 In \cite{OSV-2025}, it was shown that if the prolonged action of the pseudo-group $\G$ on the submanifold jet space $\Jn$ eventually becomes free, then the pseudo-group is reducible. Since we are interested in the construction of moving frames, which requires the prolonged action to eventually be free, in the following we can assume, without loss of generality, that $\G$ is reducible.  Also proven in \cite{OSV-2025} is the fact that when $\G$ is reducible, the involutivity of the determining equations \eqref{Gn} implies the involutivity of the reduced determining equations \eqref{rGn} at a sufficiently high order.

\begin{example}
Assuming that the Lie pseudo-group \eqref{pg1} acts on sections $s=\{(x,y,u(x,y))\}\subset \R^3$, the reduction of \eqref{pg1} to $s$ is given by
\begin{equation}\label{rpg1}
\rX = f(x),\qquad 
\rY = f_x(x)\, y + g(x),\qquad
\rU = u(x,y) + \frac{f_{xx}(x)\, y+g_x(x)}{f_x(x)}.
\end{equation}
The reduced determining equations of order $\leq 2$ are 
%
\begin{equation}\label{rG2 ex}
\begin{gathered}
\rX_y = 0,\qquad\rU = u + \frac{\rY_x}{\rX_x},\qquad \rY_y = \rX_x,\qquad
\rU_y = u_y + \frac{\rX_{xx}}{\rX_x},\\ 
\rX_{xy} = \rX_{yy} = 0,\qquad 
\rU_x = u_x + \frac{\rY_{xx}}{\rX_x}- \frac{\rY_x \rX_{xx}}{\rX_x^2},\\
\rY_{xy} = \rX_{xx},\qquad \rY_{yy} = 0,\qquad \rU_{yy} = u_{yy}.
\end{gathered}
\end{equation}
%
%
(See \cite{OSV-2023} for a detailed derivation.)
From the determining equations \eqref{rG2 ex}, we observe that $\rG\ps{2}$ can be parametrized by the parametric derivatives $\rX, \rX_x, \rX_{xx}, \rY, \rY_x, \rY_{xx}$.
\end{example}

\section{Moving frames}\label{sec: mf}

In this section, we review the classical moving frame construction for an infinite-dimensional Lie pseudo-group as presented in \cite{OP-2008} and adapted to the reducible case in  \cite{OSV-2025}.   Let $\G$ be a reducible Lie pseudo-group acting on (local) sections  $s=\{(x,u(x))\}$ of the bundle $\pi\colon M \to \mX$.  For transformations near the identity $\mathds{1}_M$, the transformed submanifold $S =\varphi(s)$ remains a section and the reduced pseudo-group $\rG$ induces an action on the $n$-th order submanifold jet space $\Jn$, given in local coordinates by
\begin{equation}\label{G action}
(\rX,\hU\n) = P\n(x,u\n,\rZ\n_{\para}),
\end{equation}
where $\rZ\n_\para$ denotes all the parametric reduced pseudo-group jets  parametrizing $\rGn$. As in \cite{OSV-2025}, we will use hats on the submanifold jet coordinates $\hU^ \alpha _J$ in order to distinguish them from the reduced diffeomorphism jet coordinates $\rU^ \alpha _A$.  The construction of a moving frame is
then accomplished by selecting a cross-section $\K\n \subset \Jn$ that is transversal to the orbits of the prolonged pseudo-group action \eqref{G action}.  As in most applications, we will assume that $\K\n$ is a coordinate cross-section defined by fixing certain components of the submanifold jet  $z\n = (x,u\n)$ to suitable constants
\[
\K\n = \big\{x^i = c^i,\; u^\alpha_J = c^\alpha_J \;|\; i, \ (\alpha;J) \in \indKn \big\},
\]
where 
\[
\indKn \subset \big\{i, \ (\alpha;J)\;|\; i\in \{1,\ldots,p\}, \; \alpha \in \{1,\ldots, q\}, \; |J| \leq n\big\}
\]
denotes the set of indices  of order $\leq n$ that determine the cross-section.  Solving the algebraic equations
\begin{equation}\label{norm eq}
P^i(x,u\n,\rZ\n_{\para}) = c^i,\qquad
P^\alpha_J(x,u\n,\rZ\n_{\para}) = c^\alpha_J,\qquad\text{with}\qquad i,\,(\alpha;J) \in \indKn,
\end{equation}
for the reduced pseudo-group parameters $\rZ\n_{\para}$ yields the (right) moving frame\footnote{To be completely accurate, the normalizations \eqref{mf formula} should then be substituted into the principal pseudo-group jets $\rZ\n_\prin$ to obtain the transformation that sends the section $s$ to its normal form $S$.}
\begin{equation}\label{mf formula}
\rZ\n_{\para} = \rrho\n(x,u\n).
\end{equation}
Transversality of the cross-section and freeness of the reduced prolonged action guarantee, via the Implicit Function Theorem, that the normalization equations \eqref{norm eq} can be locally solved for $\rZ_\para\n$ near the cross-section.  The right moving frame \eqref{mf formula} sends the function $u(x)$ to its normal form
\begin{equation}\label{nfps}
\hU^\alpha(X) = \sum_{J\,\in\, \indKa}\> \frac{c^\alpha_J}{J!}\,(X-X_0)^J + \sum_{J\not\in \indKa}\> \frac{I^\alpha_J}{J!}\,(X-X_0)^J,\qquad \alpha=1,\ldots,q,
\end{equation}
where
\[
\indKa = \big\{(\alpha; J) \in \mI_{\K}^\infty \big\}\qquad \text{with}\qquad \alpha=1,\ldots,q.
\]
In \eqref{nfps}, the constant Taylor coefficients $c^\alpha_J$ are the same as those that appear in the normalization equations \eqref{norm eq} while the Taylor coefficients $I^\alpha_J$, when expressed in terms of the submanifold jet coordinates $(x, u^{(n)})$, provide a complete set of differential invariants. We refer the reader to \cite{OSV-2025} for a general theorem that governs the convergence of the power series \eqref{nfps}.

\begin{example}
For the pseudo-group \eqref{pg1}, we refer to \cite[Example 11]{OP-2008} for the computation of the moving frame.  Similar computations based on the reduced pseudo-group action are found in \cite[Example 6.9]{OSV-2025}.
\end{example}

\section{Recursive implementation}\label{sec: rmfi}

As seen in the previous section, a moving frame sends a section to its normal form.  We will take this approach as the starting point of our recursive implementation.  To this end, consider two sections $s,S \subset M$ of the fibered manifold $\pi\colon M \to \mX$.  In local coordinates, the ``source section'' has the form $s=\{(x,u(x))\}$, while the ``target section'' is given by $S=\{(X,\hU(X))\}$. 
We will assume that the source section represents the normal form, meaning that its jet $(x,u\ii) \in \K$ lies in the cross-section, whereas the target section will be a prescribed section  that we seek to normalize via a suitable pseudo-group diffeomorphism.  In other words, we seek a diffeomorphism  $\varphi \in \G$ such that, locally, $S = \varphi (s)$, or, equivalently, $S = \varphi ^{-1}(s)$.  In terms of the reduced pseudo-group, this requires
\begin{equation}\label{nf}
\rU = \hU(\rX)\qquad \text{or, more explicitly,}\qquad U(x,u(x)) = \hU(X(x,u(x))).
\end{equation}

Equation \eqref{nf} is called the \emph{normal form equation}.  Differentiating \eqref{nf} with respect to $x$ and solving for $u\n$ provide expressions for the reduced prolonged action, as in \eqref{G action}.  In Section \ref{sec: mf} all the computations were performed at a point and, for example, the normalization equations \eqref{norm eq} were interpreted as algebraic equations for the parametric reduced pseudo-group parameters $\rZ_\para\n$.  In this section, we take a slightly different point of view.  As shown in \cite{OSV-2025}, the normal form $u(x)$ is the solution to an initial value problem.  Therefore, the equation \eqref{nf} and its derivatives hold in a neighborhood of a point, and the normalization of the pseudo-group functions should be done in a manner that they provide initial conditions for the reduced determining equations of $\rG$.  These initial conditions must be compatible with the involutivity of the determining equations. To this end, we must introduce some basic ideas from the theory of involutive systems of partial differential equations, \cite{S-2010}.    

In the following, we order the independent variables $x=(x^1,\ldots,x^p)$ using the lexicographic ordering $x^1 \prec x^2 \prec \cdots \prec x^p$.  Based on this ordering, a notion of class is defined for the reduced pseudo-group jet derivatives $\rZ^a_J$ based on that of the multi-index $J$.

\begin{definition}
The \emph{class} of a multi-index $J = (j_1,\ldots,j_n)$ is the smallest index that appears in $J$, that is
\[
\cls J = \min \big\{ j_1,\ldots, j_n \big\}.
\]
\end{definition}

\begin{remark}
The class of a derivative is not necessarily preserved under coordinate transformations, and it is necessary to work with $\delta$-regular coordinates, \cite{S-2010}.  Fortunately, $\delta$-regular coordinates are Zariski dense. See \cite{OSV-2025, S-2010} for the construction of such coordinates.
\end{remark}

The \emph{Pommaret division}, \cite{S-2010}, assigns to the derivative $\rZ^a_J$ of $\cls J = k$ the \emph{multiplicative variables} 
\begin{equation}\label{mult-ind}
\mu_k = \big\{x^1,\ldots,x^k\big\}.
\end{equation}
Successive differentiation of $\rZ^a_J$ with respect to the multiplicative variables \eqref{mult-ind} create, what is called a (Pommaret) \emph{involutive cone}:
\[
\cone(\rZ^a_J) = \big\{ \rZ^a_{J,j_{n+1},\ldots,j_\ell}\;|\; x^{j_i} \in \mu_k \big\}. 
\]

Before explaining our recursive implementation of the moving frame method based on the normal form equation \eqref{nf} we consider an example.

\begin{example}\label{ex: mf pg1}
Continuing our running example, the reduced Lie pseudo-group \eqref{rpg1} can be parametrized in terms of the parametric pseudo-group jets 
\begin{equation}\label{gpara pg1}
Z\ii_{\para} = \big\{X,Y,X_x,Y_x,X_{xx},Y_{xx}, \ldots\>\} = \big\{X,Y\} \,\cup\, \cone(X_x) \,\cup\, \cone(Y_x),
\end{equation}
containing the two involutive cones
\[
\cone(X_x) = \big\{X_{x^k}\,|\, k\geq 1\big\}\qquad\text{and}\qquad
\cone(Y_x) = \big\{ Y_{x^k}\,|\, k\geq 1\big\}.
\]
A moving frame is obtained by normalizing the parametric pseudo-group jets \eqref{gpara pg1}.

Since $Y_x(x,y) = f_{xx}(x)\,y + g_x(x)$ and
\[
\rU(x,y) = u(x,y) + \frac{f_{xx}(x)\, y + g_x(x)}{f_x(x)} = u(x,y) + \frac{Y_x(x,y)}{X_x(x)},
\]
the normal form equation \eqref{nf} is 
\begin{equation}\label{nf0 pg1}
u(x,y) + \frac{Y_x(x,y)}{X_x(x)} = \hU(X(x),Y(x,y)),
\end{equation}
where $\hU(X,Y)$ is a prescribed function and $u(x,y)$ is the normal form we aim to construct.   To simplify the notation, we have omitted the bar notation on $X$, $Y$ since $\rX(x) = X(x)$ and $\rY(x,y) = Y(x,y)$.  Using \eqref{nf0 pg1} and solving for the normal form, we obtain
\begin{equation}\label{nf pg1}
u(x,y) = \hU(X(x),Y(x,y)) - \frac{Y_x(x,y)}{X_x(x)}.
\end{equation}
In the following we set
\begin{equation}\label{Y pg1}
Y(x,y) = \tY(x) + y X_x(x).
\end{equation}
Equation \eqref{nf pg1} can be viewed as a differential equation involving the derivatives $X_x$, $Y_x$ with $y$ playing the role of a parameter.  Since the pseudo-group acts transitively on $(x,y)$, we can restrict \eqref{nf pg1} to the line $\{(x,0)\}$.  On this line, we may fix the normal form, producing the order zero normalizations
\[
x=y=0,\qquad u(x,0) = 0.
\]
The first two equations stipulate that the power series of the normal form $u(x,y)$ is based at the origin $(x,y) = (0,0)$.  Substituting the origin in the first two pseudo-group formulas in \eqref{pg1} yields
\begin{equation}\label{XY pg1}
X(0,0) = f(0) = X_0,\qquad Y(0,0) = g(0) = Y_0,
\end{equation}
where $(X_0,Y_0)$ is an arbitrary point in the plane, which corresponds to the base point of the Taylor series expansion for the prescribed function $\hU(X,Y)$.  Substituting $u(x,0)=0$ in \eqref{nf pg1} yields the normalization equation
\[
\hU(X(x),Y(x,0)) = \frac{Y_x(x,0)}{X_x(x)} = \frac{\tY_x(x)}{X_x(x)}.
\]
Solving for $\tY_x(x)$ gives the first order ordinary differential equation
\begin{equation}\label{norm1 pg1}
\tY_x(x) = \tU(x)X_x(x),\qquad \text{where} \qquad \tU(x) = \hU(X(x),\tY(x)),
\end{equation}
for $\tY(x)$ subject to the initial condition $\tY(0) = Y(0,0) = Y_0$.  Differentiating \eqref{norm1 pg1} produces formulas for the derivatives $\tY_{x^n}$ in terms of $\hU\n$ and $X_x,\ldots, X_{x^n}$.  Evaluating at the origin yields expressions for the jets $Y_{x^n}(0,0)$ of order $n \geq 1$.  Therefore, all the parametric jet variables in the involutive cone $\cone(Y_x)$ have been normalized. Together with the order zero normalizations \eqref{XY pg1}, it only remains to normalize the parametric derivatives
\[
\rZ\ii_{\para} = \cone(X_x)
\]
from \eqref{gpara pg1}.


Differentiating \eqref{Y pg1} with respect to $x$ and substituting \eqref{norm1 pg1} we obtain
\begin{equation}\label{Yx}
Y_x = \tY_x + yX_{xx} = \tU X_x + yX_{xx},
\end{equation}
where, for simplicity, we omit the dependence on $x,y$.  Substituting \eqref{Yx} into \eqref{nf pg1} gives
\begin{equation}\label{partial norm1 pg1}
u = \hU - \tU - y\frac{X_{xx}}{X_x}.
\end{equation}
Before proceeding further, we observe that working with the partially normalized normal form \eqref{partial norm1 pg1} is in stark contrast to the classical version of the moving frame algorithm, in that pseudo-group normalizations have been implemented prior to computing the prolonged action!

We now differentiate \eqref{partial norm1 pg1} with respect to $y$, noting that $\tU(x) = \hU(X(x),\tY(x))$ is a function of $x$ only, to obtain
\begin{equation}\label{uy pg1}
u_y = \hU_YX_x - \frac{X_{xx}}{X_x},
\end{equation}
Equation \eqref{uy pg1} involves the second order derivative $X_{xx}$.  As in the previous computation, we restrict \eqref{uy pg1} to the line $\{(x,0)\}$ and set $u_y(x,0)=0$ to obtain the normalization equation
\begin{equation}\label{norm2 pg1}
X_{xx}= \hU_Y(X(x),\tY(x)) X_x^2 = \tU_Y(x) X_x^2,
\end{equation}
which is a second order ordinary differential equation for $X(x)$.  The initial conditions for this equation are obtained below.  Differentiating \eqref{norm2 pg1} with respect to $x$ and evaluating the result at the origin produces formulas for the pseudo-group jets $X_{x^n}(0,0)$ of order $n\geq 2$. Thus, the parametric jet variables in the involutive cone $\cone(X_{xx})$ have now been normalized. At this stage, the only remaining unnormalized parametric pseudo-group jet variable is $\rZ_{\para}\ps{1} = \{X_x\}$ and to construct a moving frame we must normalize $X_x(0)$.

Substituting \eqref{norm2 pg1} into \eqref{partial norm1 pg1} gives the partial normal form
\begin{equation}\label{partial norm2 pg1}
u = \hU - \tU - y\, \tU_Y X_x.
\end{equation}
%
Differentiating \eqref{partial norm2 pg1} twice with respect to $y$ yields
\begin{equation}\label{uyy norm pg1}
u_{yy} = \hU_{YY}X^2_x.
\end{equation}
In contrast to the previous computations, equation \eqref{uyy norm pg1} is not restricted to coordinate axis $\{(x,0)\}$.  To normalize $X_x(0)$,
we restrict \eqref{uyy norm pg1} to the origin. We will use a zero superscript to denote such a restriction, so 
\begin{equation}\label{UJ0}
\hU_{J}^{\,0}=\hU_J(X(0),Y(0,0))=\hU_J(X_0,Y_0).
\end{equation}
Observe that $\hU_{J}^{\,0}$ is the value of the derivative at the chosen base point on the target surface.
Taking the normal form (cross-section) normalization $u_{yy}(0,0) = 1$ in \eqref{uyy norm pg1} yields the corresponding pseudo-group parameter normalization
\begin{equation}\label{Xx0}
X_x(0) = \frac{1}{\sqrt{\hU_{YY}^{\,0}}},
\end{equation}
where, for simplicity, we assume that $\hU_{YY}^{\,0} > 0$. Equation \eqref{Xx0} together with $X(0)=X_0$ provide the initial conditions for the ordinary differential equation \eqref{norm2 pg1}.  

We have now normalized all the pseudo-group parameters, \ie computed a moving frame, and it only remains to compute the first few differential invariants.  Differentiating \eqref{uyy norm pg1} with respect to $y$ yields
\begin{equation}\label{uyyy pg1}
u_{yyy} = \hU_{YYY} X_x^3.
\end{equation}
As mentioned in \eqref{nfps}, the differential invariants are given by the non-constant Taylor coefficients of the normal form.  Therefore, evaluating \eqref{uyyy pg1} at the origin, we obtain the third order differential invariant
\[
u_{yyy}(0,0) = \hU_{YYY}^{\,0} X_x^3(0) = \frac{\hU_{YYY}^{\,0}}{(\hU_{YY}^{\,0})^{3/2}}.
\]
Similarly, differentiating \eqref{uyy norm pg1} with respect to $x$ gives
\begin{align*}
u_{xyy} &= \hU_{XYY}X_x^3 + \hU_{YYY}Y_xX_x^2 + 2\, \hU_{YY}X_x X_{xx}\\
&= \big(\hU_{XYY} + \tU \hU_{YYY} + 2 \tU_Y \hU_{YY}\big)X_x^3 + y\, \hU_{YYY} \tU_Y X_x^4,
\end{align*}
where we used \eqref{Yx} and \eqref{norm2 pg1} in the second equality.  Setting $(x,y)=(0,0)$ and again using \eqref{Xx0} yields the third order differential invariant
\[
u_{xyy}(0,0) = \frac{\hU_{XYY}^{\,0} + \hU^{0}\hU_{YYY}^{\,0} 
+ 2\,\hU_{Y}^{\,0} \hU_{YY}^{\,0}}{(\hU_{YY}^{\,0})^{3/2}}. 
\]
As noted above, the zero superscripts refer to the base point of the target surface where the indicated derivatives are to be evaluated.  Since the base point is arbitrary, we can drop any explicit mention thereof and write
\[
I_{0,3} = \frac{\hU_{YYY}}{\hU_{YY}^{3/2}},\qquad I_{1,2} = \frac{\hU_{XYY} + \hU \hU_{YYY} 
+ 2\,\hU_{Y} \hU_{YY}}{\hU_{YY}^{3/2}} 
\]
for the first two differential invariants.  Also, keep in mind that the target coordinates represent a general surface, which is why the differential invariant involves capital letters, while the source coordinates refer to the normal form.
\end{example}

In Example \ref{ex: mf pg1}, we were able to recursively perform the pseudo-group normalizations \eqref{norm1 pg1}, \eqref{norm2 pg1}, \eqref{Xx0}, and compute the correct expressions for the prolonged action between each normalization.  We now explain how to implement a general recursive moving frame algorithm based on the normal form equations \eqref{nf}.

Assume $\G$ is a reducible Lie pseudo-group with involutive reduced determining equations \eqref{rGn}.  Solving equations \eqref{rGn} and their prolongation for the reduced pseudo-group derivatives will split the pseudo-group parameters $\rZ\ii$ into principal and parametric derivatives
\[
\rZ\ii = \rZ\ii_{\prin} \,\cup\, \rZ\ii_{\para}.
\]
The involutivity of the determining equations guarantees the existence of a Rees decomposition for the parametric derivatives
\begin{equation}\label{Zpara}
\rZ\ii_{\para} = \mS \ \cup \bigcup_{\rZ^a_J \,\in\, \mB} \cone(\rZ^a_J), 
\end{equation}
where $\mS$ contains a finite number of low order parametric derivatives and $\mB$ is a finite set of parametric derivatives.  Also, the involutive cones in the decomposition \eqref{Zpara} are all disjoint from each other. In \cite{S-2010} it is explained how the decomposition \eqref{Zpara} provides well-posed initial conditions for the reduced determining equations \eqref{rGn}.  For example, if one seeks a solution in the neighborhood of the origin $x=0$, then for each $\rZ^a_J \in \mB$, with multi-index of $\cls J = k$, the parametric derivatives in the cone $\cone(\rZ^a_J)$ are determined by the initial condition
\[
\rZ^a_J(x^1,\ldots,x^k,0,\ldots,0) = f^a_J(x^1,\ldots,x^k)
\]
where $f^a_J$ is a function of the multiplicative variables $\mu_k = \{x^1,\ldots,x^k\}$.  On the other hand, for every $\rZ^a_J \in \mS$ the initial condition is given by specifying the value of the function at the origin, so that
\[
\rZ^a_J(0) = c^a_J,
\]
where $c^a_J$ is a suitable constant.  The construction of a moving frame aims to normalize the parametric pseudo-group derivatives \eqref{Zpara} in such a way that it provides well-posed initial conditions to the reduced determining equations \eqref{rGn}.

To simplify the exposition, we will assume that the pseudo-group action is transitive on the base manifold $\mX$ and that it is possible to set
\begin{equation}\label{x=0}
x=0.
\end{equation}
Equation \eqref{x=0} stipulates that the power series of the normal form $u(x)$ is based at the origin. The transitivity of the pseudo-group action implies that $X \in \mS$. Substituting \eqref{x=0} into the transformation for the independent variables yields
\begin{equation}\label{Xnorm}
\rX(0) = X_0,
\end{equation}
where $X_0 = (X_0^1,\ldots,X_0^p)$ are constant values that specify the base point of the power series for the prescribed function $\hU(X)$.  Throughout the recursive moving frame algorithm, as pseudo-group parameters get normalized, these are removed from the set of unnormalized parametric derivatives \eqref{Zpara}.  This allows us to keep track of the pseudo-group parameters that remain to be normalized at the next iteration of the recursive algorithm.  In particular, with $X(0)$ normalized in \eqref{Xnorm}, we remove $X$ from $\rZ\ii_{\para}$.

Now that the initial normalization \eqref{x=0}, leading to the pseudo-group normalization \eqref{Xnorm}, has been performed, our attention moves to the normal form equation 
\eqref{nf} and its derivatives.   In the following, we introduce the difference
\begin{equation}\label{nf0}
N^\alpha(x,u,\hU,\rZ\n_{\para}) = \rU^\alpha - \hU^\alpha(\rX) = 0,\qquad \alpha=1,\ldots,q.
\end{equation}
If the pseudo-group acts trivially on the dependent variables so that $\rU^\alpha = u^\alpha$, then the normal form equations \eqref{nf0} reduce to
\begin{equation}\label{nf0 inv}
N^\alpha = u^\alpha - \hU^\alpha = 0,\qquad \alpha=1,\ldots,q,
\end{equation}
highlighting the invariance of the dependent variables.  In this particular case, the order zero normal form equations \eqref{nf0 inv} cannot be used to normalize pseudo-group parameters.  In this situation, we differentiate \eqref{nf0 inv} with respect to the independent variables $x^1,\ldots,x^p$,
\begin{equation}\label{nfpr}
\Dt_x^J N^\alpha = 0,\qquad \alpha=1,\ldots,q,
\end{equation}
until pseudo-group parameters occur.  The order zero normal form equations \eqref{nf0} are then replaced by \eqref{nfpr} in the recursive algorithm presented below.  

With that said, we now assume that the order zero normal form equations \eqref{nf0} depend explicitly on parametric pseudo-group parameters. Among the parametric derivatives appearing in \eqref{nf0}, consider those with the largest class $\cls J = k \leq p$.  From these, select a parametric derivative $\rZ^a_J$ having the highest order $|J|$ possible.  Based on this choice, all other parametric derivatives $\rZ^b_K$ occurring in \eqref{nf0} have $\cls K \leq \cls J$ and order $|K| \leq |J|$.  There are now two cases to consider depending on whether 
\begin{enumerate}[(i)]
\item $\rZ^a_J \in \mS$, or 
\item $\rZ^a_J$ is in the union of involutive cones $\displaystyle \bigcup_{\rZ^b_K \,\in\, \mB} \cone(\rZ^b_K)$.
\end{enumerate}

Starting with the first possibility, assume $\rZ^a_J \in \mS$.  In this case, select an equation $N^\alpha=0$ that contains $\rZ^a_J$ from \eqref{nf0}, and evaluate the equation at the origin to obtain the algebraic equation
\begin{equation}\label{UX0}
N^\alpha(0,u(0),\hU(X_0),\rZ\n_{\para}(0)) = 0.
\end{equation}
In \eqref{UX0}, the value of the normal form $u^\alpha(0)$ at the origin can be set equal to any suitable constant
\[
u^\alpha(0) = c^\alpha.
\] 
Equation \eqref{UX0} is then solved for $\rZ^a_J(0)$:
\begin{equation}\label{ZaJ}
\rZ^a_J(0) = F^a_J(u(0),\hU(X_0),\widehat{\rZ}{}\n_{\para}(0)),
\end{equation}
where $\widehat{\rZ}{}\n_{\para} = (\ldots, \widehat{\rZ}{}^a_J,\ldots)$ denotes the parametric derivatives of order $\leq n$ with $\rZ^a_J$ removed.  With $\rZ^a_J(0)$ normalized, we remove that derivative from the finite set of parametric derivatives $\mS$.  The normalization \eqref{ZaJ} for $\rZ^a_J(0)$ is then substituted into \eqref{nf0}.  

\begin{example}
An example where the first case discussed above occurs is  with the Lie pseudo-group
\[
X = f(x),\qquad Y = y+a,\qquad U = u+b,
\]
where $f\in \D(\R)$ and $a,b\in \R$, acting on surfaces $s=\{(x,y,u(x,y))\} \subset \R^3$.  The corresponding normal form equation is
\[
\hU(X(x),Y(y)) = u(x,y) + b.
\]
The group parameter $b$ can be normalized by setting $u(0,0) = 0$ to obtain
\[
\hU(X_0,Y_0) = b,
\]
where $X_0=X(0)$, $Y_0 = Y(0)$.
\end{example}

\begin{remark}
Any local Lie group action also falls within case (i) as there are only finitely many group parameters.  This implies that the Rees decomposition \eqref{Zpara} does not contain any involutive cones, so $\rZ\ii_\para = \mS$.
\end{remark}

We now consider case (ii) where the parametric derivative
\[
\rZ^a_J \in \bigcup_{\rZ^b_K \,\in\, \mB} \cone(\rZ^b_K)
\]
is contained in the union of disjoint involutive cones.  Involutivity implies that there exists a parametric derivative $\rZ^a_K \in \mB$ with $|K| \leq |J|$ and $\cls J \leq \cls K$, such that $\cone(\rZ^a_J) \subseteq \cone(\rZ^a_K)$.  If fact, when $|K| = |J|$ the equality $\cls J = \cls K$ must hold and $\cone(\rZ^a_J) = \cone(\rZ^a_K)$ is an involutive cone occurring in the Rees decomposition.  On the other hand, when $|K|<|J|$, then $\cls J \leq \cls K$ and $\cone(\rZ^a_J)$ is an involutive subcone strictly contained in $\cone(\rZ^a_K)$.  Indeed, the involutive cone $\cone(\rZ^a_K)$ can be decomposed into a finite set of pseudo-group parameters $\rZ^a_L$ of order $|K| \leq |L| < |J|$ and a finite collection of non-intersecting involutive cones $\cone(\rZ^a_L)$ of class $1\leq \cls L \leq k$ generated by the pseudo-group parameters $\rZ^a_L$ of order $|L| = |J|$ contained in $\cone(\rZ^a_K)$ so that
\begin{equation}\label{cone decomp}
\cone(\rZ^a_K) = \big\{\rZ^a_{L} \in \cone(\rZ^a_K)\;:\; |K| \leq |L| < |J| \big\} \;\cup\; \bigcup_{\substack{\rZ^a_{L} \,\in\, \cone(\rZ^a_K)\\ |L| = |J|}}\cone(\rZ^a_{L}),
\end{equation}
where the union contains the involutive subcone $\cone(\rZ^a_J)$.


\begin{example}
To illustrate the cone decomposition \eqref{cone decomp}, assume $x \prec y$ and $\rY=Y(x,y)$.  Then the derivatives in the cone $\cone(\rY_y)$ are given by the tree diagram
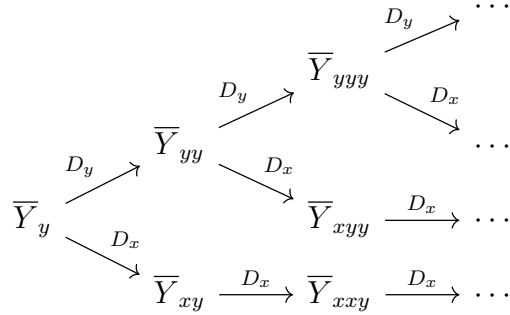
\begin{center}
\begin{tikzcd}[row sep=tiny]
& & & \cdots \\
& & \rY_{yyy} \arrow[ru, "D_y"] \arrow[rd, "D_x"] & \\
& \rY_{yy} \arrow[ru, "D_y"] \arrow[rd, "D_x"] & & \cdots\\
\rY_y \arrow[ru, "D_y"] \arrow[rd, "D_x"] & & \rY_{xyy} \arrow[r, "D_x"] & \cdots \\
& \rY_{xy} \arrow[r, "D_x"] & \rY_{xxy} \arrow[r, "D_x"] & \cdots
\end{tikzcd}
\end{center}
Therefore, we can, for example, decompose $\cone(\rY_y)$ in terms of order three subcones as follows:
\[
\cone(\rY_y) = \big\{\rY_y, \rY_{yy}, \rY_{xy} \big\} \,\cup\, \cone(\rY_{yyy})\,\cup\, \cone(\rY_{xyy}) \,\cup\, \cone(\rY_{xxy}).
\]
\end{example}

When $\cone(\rZ^a_J)$ is contained in the set of parametric pseudo-group parameters, we introduce the hyperplane $H^k=\{(x^1,\ldots,x^k,0,\ldots,0)\}$, where the nonzero components coincide with the multiplicative variables associated with the class $k$ of the multi-index $J$.  We then consider one of the normal form equations in \eqref{nf0} that contains $\rZ^a_J$ and restrict the equation to the hyperplane $H^k$ to obtain the equation
\begin{equation}\label{UHk}
N^\alpha(x,u,\hU,\rZ\n_{\para})\at{H^k} = 0
\end{equation}
for some $1 \leq \alpha \leq q$.  The restriction of the normal form component $u^\alpha$ to 
the hyperplane $H^k$ can be set equal to any suitable function of the multiplicative variables $\mu_k=\{x^1,\ldots,x^k\}$:
\begin{equation}\label{ualphaHk}
u^\alpha\at{H_k} = u^\alpha(x^1,\ldots,x^k,0,\ldots,0) = f^\alpha(x^1,\ldots,x^k).
\end{equation}
Typically, the function on the right hand side of \eqref{ualphaHk} is set equal to some constant function 
\[
u^\alpha(x^1,\ldots,x^k,0,\ldots,0) = f^\alpha(x^1,\ldots,x^k) = c^\alpha.
\]  
Solving for $\rZ^a_J$ in \eqref{UHk} yields the partial differential equation 
\begin{equation}\label{ZaJ-2}
\rZ^a_J\at{H^k} = F^a_J(x,u(x),\hU(X),\widehat{\rZ}{}\n_{\para})\at{H^k},
\end{equation}
where, as in \eqref{ZaJ}, $\widehat{\rZ}{}\n_{\para}$ denotes the unnormalized parametric derivatives of order $\leq n$ with $\rZ^a_J$ removed. The prolongation of \eqref{ZaJ-2} with respect to the multiplicative variables $\mu_k = \{x^1,\ldots,x^k\}$ then provides expressions for all the parametric derivatives in the cone $\cone(\rZ^a_J)$.  This cone is then removed from \eqref{Zpara}.  This either results in removing an entire involutive cone from the Rees decomposition \eqref{Zpara}, or removing a subcone from the cone decomposition \eqref{cone decomp}.  In the latter case, the removal of the subcone modifies the composition \eqref{Zpara} by introducing new pseudo-group parameters to $\mS$ and new subcones to the union of involutive cones.  An illustration of this situation is given in Example \ref{ex: subcone} in Section \ref{sec: ex}.  Finally, the pseudo-group normalization \eqref{ZaJ-2} is then substituted back into the normal form equation \eqref{nf0}.

\begin{example}
Example \ref{ex: mf pg1} provides an illustration of the situation just discussed above.  As we have seen, the order zero normal form equation \eqref{nf0 pg1} relates the first order pseudo-group derivatives $X_x$ and $Y_x$.  Both derivatives are of class one and of order one, and in our computations we decided to solve for $Y_x$.  Since $Y_x$ is of class one, the normal form equation \eqref{nf pg1} was then restricted to the line $H^1 = \{(x,0)\}$, producing an ordinary differential equation whose differentiation with respect to $x$ provides expressions for $Y_{x^n}\at{H^1}$.
\end{example}

Now that $\rZ^a_J$ has been normalized and substituted back into the order zero normal form equations \eqref{nf0}, we continue to perform normalizations using the remaining normal form equations $N^\beta =0$ in \eqref{nf0} with $\beta \neq \alpha$.  With $q$ equations in \eqref{nf0}, the above normalization process must be implemented at most $q$ times. Once all the equations in \eqref{nf0} have been exhausted, the first prolongation of the normal form equations is considered.  When computing the first prolongation of the normal form equation
\begin{equation}\label{Falpha}
N^\alpha(x,u,\hU,\rZ\n_{\para}) = 0
\end{equation}
used in \eqref{UX0} or \eqref{UHk} to normalize $\rZ^a_J$, we only differentiate equation \eqref{Falpha} with respect to the non-multiplicative variables $\{x^{k+1},\ldots,x^p\}$. In case (i), all the independent variables $x=(x^1,\ldots,x^p)$ are non-multiplicative variables.  Also, if an order zero normal form equation is not used during the normalization process, this indicates the presence of an order zero normalized invariant, and this equation is differentiated with respect to all the independent variables $x=(x^1,\ldots,x^p)$.  In case (ii) it is not necessary to consider prolongations with respect to the multiplicative variables $\mu_k = \{x^1,\ldots,x^k\}$ as they are used to normalize the parametric derivatives in the cone $\cone(\rZ^a_J)$.  

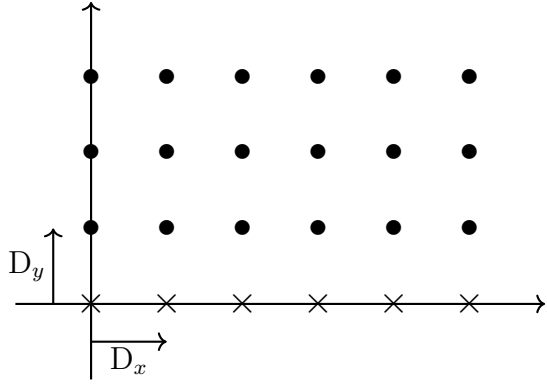
\begin{figure}
\begin{center}
\begin{tikzpicture}
    \draw[->, thick] (0,1) -- (7,1);
    \draw[->, thick] (1,0) -- (1,5);
    \foreach \i in {1, 2,..., 6} {
    \node at (\i,1) {{\large$\times$}};
    \foreach \j in {2, 3, 4} {
    \node at (\i, \j) {{\large$\bullet$}};
    }
    }
    \draw[->, thick] (1,0.5) -- (2,0.5);
    \node at (1.5,0.25) {$\Dt_x$};
    \draw[->, thick] (0.5,1) -- (0.5,2);
    \node at (0.15,1.5) {$\Dt_y$};
\end{tikzpicture}
\end{center}
\caption{Prolongation and normalization of the normal form equation $\Dt^i_x\Dt_y^j(N^\alpha)=0$. An index $(i,j)$ labeled by $\times$ represents a normal form equation used to normalize a pseudo-group parameter. An index $(i,j)$ labeled by $\bullet$ represent a normal form equation that has not yet been used to normalize a pseudo-group parameter.}\label{fig: N}
\end{figure}

Once all the appropriate order one normal form equations have been computed, we repeat the order zero normalization process that was outlined above.  Since the order one equations were obtained by differentiating the order zero equations with respect to non-multiplicative variables, any prolongation of an order one equation with respect to the multiplicative variables of a parametric pseudo-group derivative $\rZ^a_J$ during the normalization process will not intersect with previous normalizations. To illustrate this claim, assume $x\prec y$ and a normal form equation $N^\alpha=0$ (restricted to $H^1 = \{(x,0)\}$) and its prolongation $\Dt_x^i N^\alpha=0$ with respect to $x$ are used to normalized certain parametric derivatives of class one.  In Figure \ref{fig: N}, the equations $\Dt_x^iN^\alpha=0$ used to normalize the pseudo-group parameters of class one are represented by the points $(i,0)$ and are labeled by $\times$.  The prolongation of $N^\alpha$ with respect to the non-multiplicative variable $y$ yields the point $(0,1)$ in Figure \ref{fig: N}  and any point $(i,j+1)$ corresponding to the prolonged equation $\Dt_x^i\Dt_y^{j+1}(N^\alpha)=0$ can be used to normalize further pseudo-group parameters.  Note that these equations labeled by a bullet do not overlap with any previously normalized equations.

At order one, when using the normal form equation
\begin{equation}\label{N prolongation}
\Dt_{x^i} N^\alpha(x,u\ps{1},\hU\ps{1},\rZ\ps{n+1}_{\para})=0
\end{equation}
to normalize a parametric derivative, constraints on the normal form derivative $u^\alpha_i$ are to be imposed.  This originates from the fact that at the identity transformation, \eqref{N prolongation} reduces to $u^\alpha_i = \hU^\alpha_{X^i}$, and hence, near the identity transformation, the equations \eqref{N prolongation} depend on $u^\alpha_i$. Once all the order one equations have be exhausted and no further normalizations are possible, we proceed to the order two normal form equations, and so on. Assuming that the prolonged action eventually becomes free, the process of normalizing pseudo-group parameters will eventually terminate at a finite order $n$ with the creation of a moving frame.  Any remaining normal form equation 
\[
\Dt^J_x N^\alpha(x,u\ps{n},\hU\ps{n},\rrho\n(X,\hU\n))=0
\] 
not used to normalize pseudo-group parameters can be evaluated at the origin $x=0$ and then solved for $u^\alpha_J(0)$ to produce the basic differential invariant
\[
u^\alpha_J(0) = I^\alpha_J(X_0,\hU\ps{n}(X_0)).
\]
As before, the base point $X_0$ is arbitrary and hence can be omitted from the final formulae for the differential invariants $I^\alpha_J(X,\hU\ps{n})$.

\section{More examples}\label{sec: ex}

In this final section, we illustrate the recursive construction of moving frames with several additional examples.

\begin{example}
In this example, we consider the Lie pseudo-group
\[
X = f(x),\qquad Y = y,\qquad U = \frac{u}{f_x(x)} = \frac{u}{X_x},
\]
acting on surfaces $s=\{(x,y,u(x,y))\} \subset \R^3$.  This pseudo-group was used in \cite{O-2011,OV-2018} to illustrate the recursive moving frame algorithm for Lie pseudo-groups based on the universal recurrence relations, \cite{OP-2008}, and the formulas for the Maurer--Cartan forms, \cite{OP-2005}.  The reader is invited to compare those computations with the computations based on the normal form equations \eqref{nf} introduced here, and thereby confirm that the latter is more straightforward and easier to implement.

To simplify the notation, we now omit the hat notation over the prescribed surface and write $U(X,Y) = \hU(X,Y)$.   Then, the normal form equation is
\begin{equation}\label{nf eq ex1}
U(X(x),Y) = U(X(x),y) = \frac{u(x,y)}{X_x(x)},
\end{equation}
where the parametric pseudo-group parameters are
\[
Z\ii_{\para} = \{X\}\,\cup\, \cone(X_x).
\]
Assuming that the prescribed surface does not meet the $XY$-plane, \ie $U(X,Y)\neq 0$, the order zero normalizations are given by
\[
x=0\qquad \text{and}\qquad u(x,y_0) = 1,
\]
where $Y_0 = y_0$ is a constant.  The corresponding normalization equation is
\begin{equation}\label{pg norm ex1}
U(X(x),Y_0) = \frac{u(x,y_0)}{X_x(x)} = \frac{1}{X_x(x)}
\qquad\text{so that}\qquad X_x(x) = \frac{1}{U(X(x),Y_0)}.
\end{equation}
Equation \eqref{pg norm ex1} is a first order ordinary differential equation for $X(x)$ with initial condition given by 
\begin{equation}\label{X0}
X(0) = X_0,
\end{equation}
where $X_0$ is a constant. Differentiating \eqref{pg norm ex1} with respect to $x$ and evaluating the result at $x=0$ produces, along with the initial condition \eqref{X0}, formulas for the pseudo-group jets $X_{x^n}(0)$ for all $n\geq 0$.  Therefore, we have finished constructing a moving frame. Substituting the pseudo-group normalization \eqref{pg norm ex1} into \eqref{nf eq ex1} yields
\begin{equation}\label{inv ex1}
U(X(x),Y) = u(x,y)\, U(X(x),Y_0)\qquad\text{or}\qquad
u(x,y) = \frac{U(X(x),Y)}{U(X(x),Y_0)}.
\end{equation}
Differentiating \eqref{inv ex1} with respect to $y$ gives
\begin{equation}\label{norm u ex1}
u_y(x,y) = \frac{U_Y(X(x),Y)}{U(X(x),Y_0)},
\end{equation}
as $Y=y$. At $(x,y) = (0,y_0)$, the expression \eqref{norm u ex1} produces the differential invariant
\[
u_y(0,y_0) = \frac{U_Y(X_0,Y_0)}{U(X_0,Y_0)} = \frac{U_{Y}^{\,0}}{U^0},
\]
where, as before, $U_{J}^{\,0} = U_J(X_0,Y_0)$ refers to the value of the derivative at the base point of the target surface.  Differentiating \eqref{norm u ex1} with respect to $x$ gives
\[
u_{xy}(x,y) = \frac{U_{XY}(X(x),Y) X_x(x)}{U(X(x),Y_0)} - \frac{U_Y(X(x),Y) U_X(X(x),Y_0) X_x(x)}{U(X(x),Y_0)^2}.
\]
Substituting the pseudo-group normalization \eqref{pg norm ex1} yields
\[
u_{xy}(x,y) = \frac{U_{XY}(X(x),Y)}{U(X(x),Y_0)^2} - \frac{U_Y(X(x),Y) U_X(X(x),Y_0)}{U(X(x),Y_0)^3}.
\]
Setting $(x,y) = (0,y_0)$, we obtain the second order differential invariant
\[
u_{xy}(0,y_0) = \frac{U^0 U_{XY}^{\,0} - U_{X}^{\,0} U_{Y}^{\,0}}{(U^0)^3}.
\]
Differentiating \eqref{norm u ex1} with respect to $y$ gives
\[
u_{yy}(x,y) = \frac{U_{YY}(X(x),Y)}{U(X(x),Y_0)},
\]
which produces the second order differential invariant
\[
u_{yy}(0,y_0) = \frac{U_{YY}^{\,0}}{U^0}.
\]
Again, we can drop the explicit reference to the arbitrary base point and obtain the differential invariants
\[
I_{0,1} = \frac{U_Y}{U},\qquad 
I_{1,1} = \frac{U U_{XY} - U_X U_Y}{U^3},\qquad
I_{0,2} = \frac{U_{YY}}{U}.
\]  
Higher order differential invariants can be obtained either through further application of the preceding differentiation process or, alternatively, by using invariant differentiation and the recurrence formulas, as in \cite{OP-2008}.  
\end{example}

\begin{example}
In this  example, we consider the Lie pseudo-group
\[
X = f(x),\qquad Y = g(x,y),\qquad Z = z+a,\qquad U=u+b,
\]
where $f_x\, g_y \neq 0$ and $a,b \in \mathbb{R}$, acting on hypersurfaces given by the graph of $u = u(x,y,z)$.  The order one reduced determining equations are 
\begin{equation}\label{rdet pg3}
\begin{gathered}
\rX_y = \rX_z = 0,\qquad \rY_z=0,\qquad \rZ_x = \rZ_y = 0,\quad \rZ_z=1,\\
\rU_x = u_x,\quad \rU_y = u_y,\quad \rU_z = u_z.
\end{gathered}
\end{equation}
Using the ordering $x \prec y \prec z$, the equations \eqref{rdet pg3} are involutive with the set of parametric derivatives given by
\begin{equation}\label{Rees pg3}
Z_{\para}\ii = \{X,Y,Z,U\} \,\cup\, \cone(X_x) \,\cup\, \cone(Y_x) \,\cup\, \cone(Y_y),
\end{equation}
where we dropped the bar notation.  In contrast to the previous examples where the parametric pseudo-group parameters only contained involutive cones of class one, the Rees decomposition \eqref{Rees pg3} includes the involutive cone $\cone(Y_y)$ of class two.

In the following we work on the set of regular jets
\[
\mV^2 = \big\{ u_y \neq 0, \quad u_{xz}u_y - u_{yz} u_x \neq 0\big\}.
\]
The normal form equation is
\begin{equation}\label{nf pg3}
U(X,Y,Z) = u(x,y,z) + b.
\end{equation}
The order zero normalizations are
\[
x=y=z=u(0,0,0)=0,
\]
producing the pseudo-group normalizations
\begin{gather*}
X(0) =f(0) = X_0,\qquad Y(0,0) = g(0,0) = Y_0,\qquad Z(0) = a = Z_0,\\ U(X_0,Y_0,Z_0) = b = U_0,
\end{gather*}
where $X_0$, $Y_0$, $Z_0,U_0$ determine the base point on the prescibed hypersurface.  Differentiating the normal form equation \eqref{nf pg3} with respect to $x,y,z$ yields 
\begin{equation}\label{nf1 pg3}
U_X X_x + U_Y Y_x = u_x,\qquad 
U_Y Y_y = u_y,\qquad U_Z = u_z,
\end{equation}
where the third equation shows that $I_{0,0,1} = U_Z$ is a differential invariant. Starting with the second equation in \eqref{nf1 pg3}, since $Y_y$ is of class two according to the ordering $x\prec y \prec z$, we can set $u_y(x,y,0) = 1$ producing the normalization
\begin{equation}\label{Yy}
Y_y(x,y) = \frac{u_y(x,y,0)}{U_Y(X(x),Y(x,y),Z_0)} 
= \frac{1}{U_Y(X(x),Y(x,y),Z_0)}.
\end{equation}
The normalization equation \eqref{Yy} is a first order partial differential equation, which, when differentiated with respect to $x, y$ and evaluated at the origin, produces formulas for the pseudo-group jets $Y_{x^j,y^{k+1}}(0,0)$ contained in the involutive cone $\cone(Y_y)$. 

We now consider the first equation in \eqref{nf1 pg3}.  Under our assumption that $U_Y \neq 0$ we can solve the equation for $Y_x$.  Since $Y_x$ is of class one, we can set $u_x(x,0,0) = 0$ to obtain
\begin{equation}\label{Yx pg3}
Y_x(x,0) = -\frac{U_X(X(x),Y(x,0),Z_0)X_x(x)}{U_Y(X(x),Y(x,0),Z_0)} = -\frac{\tU_XX_x(x)}{\tU_Y},
\end{equation}
where we abbreviate $\tU_J = U_J(X(x),Y(x,0),Z_0)$.  Equation \eqref{Yx pg3} is a first order ordinary differential equation which, when differentiated with respect to $x$ and then evaluated at the origin, yields expressions for the remaining pseudo-group jets $Y_{x^{j+1}}(0,0)$ forming the involutive cone $\cone(Y_x)$.

Differentiating the third equation in \eqref{nf1 pg3} we obtain
\begin{equation}\label{nf2 pg3}
U_{XZ} X_x + U_{YZ} Y_x = u_{xz},\qquad
U_{YZ} Y_y = u_{yz},\qquad
U_{ZZ} = u_{zz}.
\end{equation}
Restricting the first equation in \eqref{nf2 pg3} to the line $\{(x,0,0)\}$ and substituting \eqref{Yx pg3}, we obtain the equation
\[
\frac{(\tU_{XZ}\tU_Y - \tU_{YZ}\tU_X)X_x(x)}{\tU_Y} = u_{xz}(x,0,0).
\]
Setting $u_{xz}(x,0,0)=1$ yields the first order ordinary differential equation
\[
X_x(x) = \frac{\tU_Y}{\tU_{XZ}\tU_Y - \tU_{YZ}\tU_X},
\]
thereby, after prolongation, producing formulas for the pseudo-group derivatives $X_{x^{j+1}}$ contained in the involutive cone $\cone(X_x)$. At this stage, all parametric pseudo-group parameters have been normalized and a moving frame has been constructed. Evaluating the remaining equations in \eqref{nf2 pg3} at the origin produces the differential invariants
\[
u_{yz}(0,0,0) = \frac{U_{YZ}^{\,0}}{U_{Y}^{\,0}},\qquad 
u_{zz}(0,0,0) = U_{ZZ}^{\,0},
\]
where $U_{J}^{\,0} = U_J(X_0,Y_0,Z_0)$.  As usual, higher order differential invariants can be obtained by invariant differentiation using the recurrence formulae or by further differentiating the normal form equation.
\end{example}

\begin{example}
All previously considered examples have involved Lie pseudo-group actions that were quasi-horizontal, as defined in \cite{A-2022}, in the chosen system of coordinates.  This property is not necessary for the recursive implementation introduced in this paper, and we illustrate this fact by considering the Lie pseudo-group 
\begin{equation}\label{pg4}
X = x+a,\qquad Y=y+b,\qquad U = f(u),
\end{equation}
where $a,b \in \R$ and $f\in \D(\R)$.  Of course, the Lie pseudo-group \eqref{pg4} can be transformed into a quasi-horizontal action via the hodograph transformation $(x,y,u) \to (u,y,x)$, but we will not make this transformation here.  

Assuming $u = u(x,y)$, the reduced determining equations are
\begin{equation}\label{rdeteq pg4}
X_x=1,\quad X_y = 0,\qquad 
Y_x=0,\quad Y_y = 1,\qquad 
u_y U_x = u_x U_y,
\end{equation}
where we omit the bar notation to denote the reduced pseudo-group parameters. The determining equations \eqref{rdeteq pg4} are involutive using the ordering $x \prec y$.  The set of parametric pseudo-group derivatives is given by
\[
Z\ii_{\para} = \{X,Y,U\}\,\cup\, \cone(U_x).
\]
We now construct a moving frame recursively on the set of regular jets $\nabla u = (u_x,u_y)\neq 0$, restricting our computations to the subset
\[
\mV\ii = \{u_x\neq 0\}.
\]
The normal form equation for this example is
\begin{equation}\label{nf pg4}
\hU(X,Y) = U.
\end{equation}
At order zero, we set
\[
x=0,\qquad y=0,\qquad u(0,0)=0.
\]
This normalizes the parametric pseudo-group parameters $\{X,Y,U\}$:
\[
X(0) = X_0,\qquad Y(0)=Y_0,\qquad U(0) = \hU^{0}
\]
where $\hU^{0} = \hU(X_0,Y_0)$.  Differentiating \eqref{nf pg4} gives
\begin{equation}\label{nfpr pg4}
\hU_X = U_x,\qquad \hU_Y = U_y = \frac{u_y U_x}{u_x}.
\end{equation}
Using the first equation in \eqref{nfpr pg4}, we can set
\begin{equation}\label{ux pg4}
u_x(x,0)=1
\end{equation}
to obtain the pseudo-group normalization equation
\begin{equation}\label{mf pg4}
U_x(x,0) = \hU_X(X(x),Y_0).
\end{equation}
Differentiating \eqref{mf pg4} with respect to $x$ and evaluating the result at the origin provides formulas for the parametric pseudo-group jets $U_{x^{j+1}}$ forming the involutive cone $\cone(U_x)$. Evaluating the second equation in \eqref{nfpr pg4} at the origin yields the differential invariant
\[
u_y(0,0) = \frac{\hU_{Y}^{\,0}}{\hU_{X}^{\,0}}.
\]
We obtain higher order differential invariants by differentiating the second equation in \eqref{nfpr pg4}.  First, differentiation with respect to $x$ gives 
\begin{equation}\label{uxy pg4}
\hU_{XY} = \frac{u_xu_y U_{xx}+ (u_xu_{xy} - u_y u_{xx}) U_x}{u_x^2}.
\end{equation}
Differentiating \eqref{ux pg4} implies $u_{xx}(x,0) = 0$.  Similarly, differentiating \eqref{mf pg4} gives $U_{xx}(0,0) = \hU_{XX}(X_0,Y_0)$.  Thus, evaluating \eqref{uxy pg4} at the origin produces the differential invariant  
\[
u_{xy}(0,0) = \frac{\hU_X^{\,0}\hU_{XY}^{\,0} - \hU_{Y}^{\,0}\hU_{XX}^{\,0}}{(\hU_{X}^{\,0})^2}.
\]
Next, differentiating the second equation in \eqref{nfpr pg4} with respect to $y$ gives
\begin{equation}\label{UYY pg4}
\hU_{YY} = \frac{u_xu_y U_{xy} + (u_xu_{yy} - u_y u_{xy})U_x}{u_x^2}.
\end{equation}
Differentiating the last determining equation in \eqref{rdeteq pg4} with respect to $x$ we obtain
\begin{equation}\label{Uxy pg4}
U_{xy} = \frac{u_xu_yU_{xx} + (u_x u_{xy} - u_y u_{xx}) U_x}{u_x^2}.
\end{equation}
Substituting \eqref{Uxy pg4} in \eqref{UYY pg4} gives
\[
\hU_{YY} = \frac{u_xu_y^2 U_{xx} + (u_x^2u_{yy} - u_y^2 u_{xx})U_x }{u_x^3}.
\]
Evaluating the result at the origin produces the differential invariant
\[
u_{yy}(0,0) = \frac{(\hU_{X}^{\,0})^2\hU_{YY}^{\,0} - (\hU_{Y}^{\,0})^2 \hU_{XX}^{\,0}}{(\hU_{X}^{\,0})^3}.
\]
\end{example}

\begin{example}\label{ex: subcone}
In the previous examples, the recursive moving frame algorithm always led to the normalization of entire involutive cones in the Rees decomposition of the parametric reduced pseudo-group parameters.  In this example, we consider a pseudo-group where the recursive moving frame algorithm normalizes a subcone.  To this end, consider the Lie pseudo-group
\begin{equation}\label{pg6}
X = x+a,\qquad Y = f(x,y),\qquad Z = z + b,\qquad U = u + f_{yy}(x,y),
\end{equation}
where $f_y \neq 0$ and $a,b \in \mathbb{R}$ acting on hypersurfaces given by $u = u(x,y,z)$.  As in previous examples, we omit the bar notation to denote reduced pseudo-group parameters. We use the ordering $x \prec y \prec z$ so that the parametric reduced pseudo-group parameters are given by the Rees decomposition
\begin{equation}\label{pg para}
\{X,Y,Z\} \,\cup\, \cone(Y_x) \,\cup\, \cone(Y_y).
\end{equation}

Implementing the recursive moving frame algorithm, we first normalize
\[
x=y=z=0,
\]
so that
\begin{equation}\label{XYZ}
X(0) = X_0,\qquad Y(0,0) = Y_0,\qquad Z(0) = Z_0.
\end{equation}
The first and last normalizations in \eqref{XYZ} are equivalent to normalizing $a=X_0$, $b=Z_0$ in the pseudo-group expression \eqref{pg6}.  Next, we consider the normal form equation
\begin{equation}\label{nf6}
\hU(X(x),Y(x,y),Z(z)) = u(x,y,z) + Y_{yy}(x,y).
\end{equation}
With $\cls\!(Y_{yy}) = 2$, we restrict \eqref{nf6} to the coordinate plane $\{(x,y,0)\}$ and set $u(x,y,0)=0$ to obtain the partial differential equation
\begin{equation}\label{Yyy}
Y_{yy}(x,y) = \hU(X(x),Y(x,y),Z_0),
\end{equation}
where $X(x)$ is the solution to the initial value problem
\[
X_x(x) = 1,\qquad X(0) = X_0\qquad\text{so that}\qquad X(x) = x+X_0.
\]
Differentiating \eqref{Yyy} with respect to the multiplicative variables $x$, $y$ provides formulas for the parametric derivatives in the involutive cone $\cone(Y_{yy})$, which is a subcone of
\[
\cone(Y_y) = \{Y_y\} \,\cup\, \cone(Y_{xy}) \,\cup\, \cone(Y_{yy}).
\]
Therefore, after performing the order zero normalizations \eqref{XYZ}, \eqref{Yyy}, the pseudo-group parameters from \eqref{pg para} that remain to be normalized are
\begin{equation}\label{Rees6 pnorm}
\{Y_y\} \,\cup\, \cone(Y_x) \,\cup\, \cone (Y_{xy}).
\end{equation}
Substituting \eqref{Yyy} into the normal form equation \eqref{nf6} yields
\begin{equation}\label{nf6 pnorm}
\hU(X(x),Y(x,y),Z(z)) = u(x,y,z) + \hU(X(x),Y(x,y),Z_0).
\end{equation}
Differentiating \eqref{nf6 pnorm} with respect to the non-multiplicative variable $z$, and using the fact that $Z_z = 1$, we obtain the differential invariant
\begin{equation}\label{uz}
\hU_Z = u_z.
\end{equation}
Differentiating \eqref{uz} with respect to $x, y, z$ yields
\begin{equation}\label{pr2}
\hU_{XZ} + \hU_{YZ} Y_x = u_{xz},\qquad \hU_{YZ} Y_y = u_{yz},\qquad
\hU_{ZZ} = u_{zz}.
\end{equation}
Assuming $\hU_{YZ} \neq 0$, we restrict the first equation in \eqref{pr2} to the coordinate axis $\{(x,0,0)\}$ and set $u_{xz}(x,0,0) = 0$ to obtain the first order ordinary differential equation
\begin{equation}\label{Yx6}
Y_x(x,0) = -\frac{\tU_{XZ}}{\tU_{YZ}},
\end{equation}
where the tilde notation denotes the restriction to the line $\{(x,0,0)\}$ so that, for example, $\tU_{XZ} = \hU_{XZ}(X(x),Y(x,0),Z_0)$.  Differentiating equation \eqref{Yx6} provides formulas for the parametric derivatives in the involutive cone $\cone(Y_x)$.  

Next, since the parametric pseudo-group parameter $Y_y$ in \eqref{Rees6 pnorm} is not contained in an involutive cone, its normalization is achieved by evaluating the second equation in \eqref{pr2} at the origin and setting $u_{yz}(0,0,0)=1$.  This produces the algebraic equation
\[
Y_y(0,0) = \frac{1}{\hU^0_{YZ}},
\]
where $\hU^0_{YZ} = \hU_{YZ}(X_0,Y_0,Z_0)$.  At this stage, it only remains to normalize the pseudo-group parameters in the involutive cone
\[
\cone(Y_{xy}).
\]
To do so, we differentiate the first equation in \eqref{pr2} with respect to the non-multiplicative variable $y$ to obtain
\[
\hU_{XYZ} Y_y + \hU_{YYZ} Y_y Y_x + \hU_{YZ} Y_{xy} = u_{xyz}.
\]
Restricting this equation to the line $\{(x,0,0)\}$ and setting $u_{xyz}(x,0,0)=0$, we obtain
\[
Y_{xy}(x,0) = Y_y(x,0) \bigg(\frac{\tU_{YYZ}\tU_{XZ} - \tU_{XYZ}\tU_{YZ}}{\tU_{YZ}^2} \bigg),
\]
where, using the pseudo-group normalization \eqref{Yyy}, 
\[
Y_y(x,0) = Y_y(0,0) + \int_0^x Y_{yy}(t,0)\, \dt t 
= \frac{1}{\hU^0_{YZ}} + \int_0^x \hU(X(t),Y(t,0),Z_0)\, \dt t.
\]
This ends the normalization process and the construction of a moving frame.  
\end{example}

\begin{example}
The ideas developed here also apply to finite-dimensional Lie group actions.  As an example, we consider the group of linear factional transformations
\[
X = x,\qquad U = \frac{\alpha u + \beta}{\gamma u + \delta},
\]
where $\alpha\delta - \beta\gamma =1$.  Assuming $u=u(x)$, the reduced determining equation is
\begin{equation}\label{ex3 det eq}
\rU_{xxx} = \frac{3}{2} \frac{\rU_{xx}}{\rU_x} - \frac{3}{2}\frac{u_{xx}^2\rU_x}{u_x^2} + \frac{u_{xxx}\rU_x }{u_x}.
\end{equation}
Therefore, the parametric pseudo-group derivatives are
\[
\rU,\qquad \rU_x,\qquad \rU_{xx}.
\]
Differentiating the normal form equation $U(X) = \rU$ we obtain
\begin{equation}\label{frac nfpr}
\begin{aligned}
&U_X = \rU_x,\qquad U_{XX} = \rU_{xx},\\ 
&U_{XXX} = \rU_{xxx} = \frac{3}{2} \frac{\rU_{xx}}{\rU_x} - \frac{3}{2}\frac{u_{xx}^2\rU_x}{u_x^2} + \frac{u_{xxx}\rU_x}{u_x}=\frac{3}{2} \frac{U_{XX}}{U_X} - \frac{3}{2}\frac{u_{xx}^2U_X}{u_x^2} + \frac{u_{xxx}U_X }{u_x},
\end{aligned}
\end{equation}
where in the last equation we used \eqref{ex3 det eq}.  Since the action is not transitive on the independent variable, we perform the computations at the point $X_0=x_0$. Then, normalizations of the normal form are given by
\[
u(x_0)=0,\qquad u_x(x_0)=1,\qquad u_{xx}(x_0)=0,
\]
which yields the pseudo-group normalizations
\[
U(X_0) = \rU(x_0),\qquad 
U_X(X_0) = \rU_x(x_0),\qquad
U_{XX}(X_0) = \rU_{xx}(x_0).
\]
Evaluating the last equation in \eqref{frac nfpr} at the base point $X_0=x_0$, with $U_{J}^{\,0} = U_J(X_0)$, we recover the Schwarzian derivative:
\[
u_{xxx}(x_0) = \frac{U_{XXX}^{\,0}}{U_{X}^{\,0}} - \frac{3}{2}\frac{(U_{XX}^{\,0})^2}{(U_{X}^{\,0})^2}.
\]
As before, since the base point is arbitrary, the zero superscript can be dropped when writing out this basic differential invariant.  Since we are dealing with a finite-dimensional Lie group, all higher order differential invariants can easily be obtained by differentiation with respect to the invariant differential operator obtained by invariantizing the operator of implicit differentiation using the moving frame, \cite{O-2012}.
\end{example}

\end{document}